\documentclass{llncs}
\usepackage{amsmath,latexsym,amssymb}

\usepackage{upgreek}
\usepackage[
  pdftitle={A family of abstract interpretations for static analysis of concurrent higher-order programs},
  pdfauthor={Matthew Might and David Van Horn},
  pdfsubject={Computer Science},
  pdfkeywords={abstract interpretation, concurrency, parallelism, k-CFA},
  breaklinks=true
]
{hyperref}

\newcommand{\mylongtitle}{A family of abstract interpretations for static analysis of concurrent higher-order programs}

\newcommand{\myshorttitle}{Short title}

\newcommand{\eg}{\emph{e.g.}}

\newcommand{\syn}[1]{\mathsf{#1}}
\newcommand{\var}[1]{\mathit{#1}}
\newcommand{\s}[1]{\mathit{#1}}

\newcommand{\parto}{\rightharpoonup}
\newcommand{\dom}{\var{dom}}

\newcommand{\set}[1]{\left\{#1\right\}}
\newcommand{\setbuild}[2]{\left\{ #1 : #2\right\}}

\newcommand{\Pow}[1]{{\mathcal{P}\left(#1\right)}}
\newcommand{\PowSm}[1]{{\mathcal{P}(#1)}}

\newcommand{\union}{\cup}

\newcommand{\To}{\mathrel{\Rightarrow}}

\newcommand{\wt}{\sqsubseteq}

\newcommand{\join}{\sqcup}

\DeclareMathOperator{\lfp}{lfp}

\newcommand{\longand}{\;\;\;\;\;\;\;}
\newcommand{\infer}[2]{{\renewcommand{\arraystretch}{1.4}\begin{array}{c}#1\\ \hline #2\end{array}\renewcommand{\arraystretch}{1.0}}}

\newcommand{\sembr}[1]{\ensuremath{[\![{#1}]\!]}}

\newcommand{\opor}{\mathrel{|}}

\newcommand{\produces}{\mathrel{::=}}

\newcommand{\vv}{v}

\newcommand{\lam}{\ensuremath{\var{lam}}}

\newcommand{\call}{\ensuremath{\var{call}}}

\newcommand{\schfalse}{{\mbox{\tt\#f}}}

\newcommand{\ttlp}{\mbox{\tt (}}
\newcommand{\ttrp}{\mbox{\tt )}}

\newcommand{\appform}[2]{\ttlp #1\; #2\ttrp}
\newcommand{\lamform}[2]{\ttlp \uplambda\;\ttlp#1\ttrp\;#2\ttrp}

\newcommand{\ifform}[3]{\ttlp {\tt if}\; #1\; #2\; #3\ttrp}

\newcommand{\setbform}[2]{\ttlp \mbox{\tt set!}\; #1\; #2 \ttrp}

\newcommand{\letiform}[3]{\ttlp {\tt let}\; \ttlp\ttlp#1\; #2\ttrp\ttrp\; #3\ttrp}

\newcommand{\fexpr}{f}
\newcommand{\expr}{e}
\newcommand{\aexpr}{\mbox{\sl {\ae}}}

\newcommand{\Eval}{{\mathcal{E}}}

\newcommand{\Inject}{{\mathcal{I}}}

\newcommand{\State}{\Sigma}
\newcommand{\state}{\varsigma}

\newcommand{\Env}{\s{Env}}

\newcommand{\Den}{D} 

\newcommand{\den}{d}

\newcommand{\store}{\sigma}

\newcommand{\env}{\rho}

\newcommand{\clo}{\var{clo}}
\newcommand{\cont}{\kappa}

\newcommand{\alloc}{\mathit{alloc}}

\newcommand{\addr}{a}

\newcommand{\val}{\var{val}}

\newcommand{\aTo}{\leadsto}

\newcommand{\aInject}{{\hat{\mathcal{I}}}}

\newcommand{\sa}[1]{\widehat{\mathit{#1}}}

\newcommand{\aEval}{{\hat{\mathcal{E}}}}

\newcommand{\atf}{{\hat{f}}}

\newcommand{\aMState}{{\hat{\Xi}}}
\newcommand{\amstate}{{\hat{\xi}}}

\newcommand{\aState}{{\hat{\Sigma}}}
\newcommand{\astate}{{\hat{\varsigma}}}

\newcommand{\aDen}{\hat{D}} 

\newcommand{\astore}{{\hat{\sigma}}}
\newcommand{\aenv}{{\hat{\rho}}}

\newcommand{\aclo}{{\widehat{\var{clo}}}}

\newcommand{\acont}{{\hat{\kappa}}}

\newcommand{\aden}{{\hat{d}}}

\newcommand{\aaddr}{{\hat{\addr}}}
\newcommand{\aalloc}{{\widehat{alloc}}}

\newcommand{\aval}{{\widehat{\var{val}}}}

\newcommand{\absmap}{\alpha}
\newcommand{\abs}[1]{|#1|}

\begin{document}
\title{\mylongtitle}
\titlerunning{\myshorttitle}
\author{Matthew Might \and David Van Horn}
\authorrunning{Matthew Might \and David Van Horn}   
%

\institute{
University of Utah and Northeastern University \\
\email{might@cs.utah.edu} and \email{dvanhorn@ccs.neu.edu} \\
\texttt{http://matt.might.net/} and \texttt{http://lambda-calcul.us/}
}

\maketitle              

\begin{abstract}
  We develop a framework for computing two foundational analyses for
\emph{concurrent} higher-order programs: (control-)flow analysis (CFA) and
may-happen-in-parallel analysis (MHP).  We pay special attention to the unique
challenges posed by the unrestricted mixture of first-class continuations and
dynamically spawned threads.  To set the stage, we formulate a concrete model
of concurrent higher-order programs: the P(CEK*)S machine.  We find that the
systematic abstract interpretation of this machine is capable of computing both
flow and MHP analyses.  Yet, a closer examination finds that the precision for
MHP is poor.  As a remedy, we adapt a shape analytic
technique---singleton abstraction---to dynamically spawned threads (as opposed
to objects in the heap).  We then show that if MHP analysis is not of interest,
we can substantially accelerate the computation of flow analysis alone
by collapsing thread interleavings with a second layer of abstraction.

\end{abstract}

\newcommand{\PCEKS}{P(CEK$^\star$)S}

\newcommand{\setsm}[1]{\{#1\}}

\newcommand{\ctxt}{c}
\newcommand{\actxt}{{\hat c}}

\newcommand{\tid}{t}
\newcommand{\atid}{{\hat{t}}}

\newcommand{\threads}{T} \newcommand{\athreads}{{\hat T}}

\newcommand{\ato}{\multimap}

\newcommand{\atcount}{{\hat \mu}}

\renewcommand{\call}{\mathit{cexp}}

\newcommand{\hist}{h}
\newcommand{\ahist}{{\hat h}}

\newcommand{\record}{\mathit{record}}
\newcommand{\arecord}{{\widehat{\mathit{record}}}}

\newcommand{\newtid}{\mathit{newtid}}
\newcommand{\anewtid}{{\widehat{\mathit{newtid}}}}

\newcommand{\aReachable}{{\hat{\mathcal{R}}}}

\newcommand{\ID}{id}
\newcommand{\IDs}{ids}

\section{Higher-order is hard; concurrency makes it harder}

\noindent 
\emph{The next frontier in static reasoning for higher-order programs is
concurrency.}
When unrestricted concurrency and higher-order computation meet, their challenges to
static reasoning reinforce and amplify one another.

Consider the possibilities opened by a mixture of dynamically created threads
and first-class continuations.
Both pose obstacles to static analysis by themselves, yet the challenge of
reasoning about a continuation created in one thread and invoked in another is
substantially more difficult than the sum of the individual challenges.

We respond to the challenge by (1) constructing the \PCEKS{} machine, a
nondeterministic abstract machine that concretely and fully models
higher-orderness and concurrency; and then (2) systematically deriving 
abstract interpretations of this machine to enable the sound and meaningful
flow analysis of concurrent higher-order programs.

Our first abstract interpretation creates a dual hierarchy of flow and
may-happen-in-parallel (MHP) analyses parameterized by context-sensitivity and
the granularity of an abstract partition among threads.
The context-sensitivity knob tunes flow-precision as in Shivers's
$k$-CFA~\cite{mattmight:Shivers:1991:CFA}.  
The partition among threads tunes the precision of MHP analysis, since it
controls the mapping of concrete threads onto abstract threads.
To improve the precision of MHP analysis, our second abstract interpretation
introduces shape analytic concepts---chiefly, singleton cardinality
analysis---but it applies them to discover the ``shape'' of threads rather than
the shape of objects in the heap.
The final abstract interpretation accelerates the computation of flow analysis
(at the cost of MHP analysis) by inflicting a second abstraction that soundly
collapses all thread interleavings together.


\subsection{Challenges to reasoning about higher-order concurrency}

The combination of higher-order computation and concurrency introduces design
patterns that challenge conventional static reasoning techniques.

\paragraph{Challenge: Optimizing futures}
Futures are a popular means of enabling parallelism in functional programming.
Expressions marked {\tt future} are computed in parallel with their own
continuation.
When that value reaches a point of strict evaluation, the thread of the
continuation joins with the thread of the future.

Unfortunately, the standard implementation of futures~\cite{mattmight:Feeley:1993:Future} inflicts
substantial costs on sequential performance: that implementation transforms
\texttt{(future $e$)} into \texttt{(spawn $e$)}, and all strict expressions
into conditionals and thread-joins.
That is, if the expression $e'$ is in a strict evaluation position, then it
becomes: \begin{center} \texttt{(let ([\$t $e'$]) (if (thread? \$t) (join \$t)
\$t))} \end{center} Incurring this check at all strict points is costly.
A flow analysis that works for concurrent programs would find that most
expressions can never evaluate to future value, and thus, need not incur such
tests.

\paragraph{Challenge: Thread cloning/replication}
The higher-order primitive {\tt call/cc} captures the current
continuation and passes it as a first-class value to its argument.
The primitive {\tt call/cc} is extremely powerful---a brief interaction between \texttt{spawn} and \texttt{call/cc} 
effortlessly expresses thread replication:
\begin{verbatim} 
           (call/cc (lambda (cc) (spawn (cc #t)) #f))\end{verbatim}
This code captures the current continuation, spawns a new thread and replicates
the spawning thread in the spawned thread by invoking that continuation.
The two threads can be distinguished by the return value of \texttt{call/cc}:
the replicant returns true and the original returns false.

\paragraph{Challenge: Thread metamorphosis}
Consider a web server in which continuations are used to suspend and restore
computations during interactions with the client~\cite{dvanhorn:Queinnec2004Continuations}.
Threads ``morph'' from one kind of thread (an interaction thread or a worker
thread)  to another by invoking continuations.
The \texttt{begin-worker} continuation metamorphizes
the calling thread into a worker thread:
\begin{verbatim}
 (define become-worker
  (let ([cc (call/cc (lambda (cc) (cc cc)))])
   (cond 
    [(continuation? cc)       cc]
    [else                     (handle-next-request)
                              (become-worker #t)])))\end{verbatim}
The procedure \texttt{handle-next-request} checks
whether the request is the resumption of an old session, and if so,
invokes the continuation of that old session:
\begin{verbatim}
 (define (handle-next-request)
  (define request (next-request))
  (atomic-hash-remove! (session-id request)
   (lambda (session-continuation) 
     (define answer (request->answer request))
     (session-continuation answer))
   (lambda () (start-new-session request)))) \end{verbatim}
When a client-handling thread needs data from the client,
it calls \texttt{read-from-client}, it associates
the current continuation to the active session,
piggy-backs a request to the client on an outstanding reply
and the metamorphizes into a worker thread to handle
other incoming clients:
\begin{verbatim}
 (define (read-from-client session)
  (call/cc (lambda (cc)
   (atomic-hash-set! sessions (session-id session) cc)
   (reply-to session))
   (become-worker #t)))\end{verbatim}

%
%

\section{\PCEKS{}: An abstract machine model of concurrent, higher-order computation}

In this section, we define a \PCEKS{} machine---a CESK machine with a pointer
refinement that allows concurrent threads of execution.
It is directly inspired by the sequential abstract machines in Van Horn and Might's
recent work~\cite{dvanhorn:VanHorn2010Abstracting}.
Abstract interpretations of this machine perform
both flow and MHP analysis for concurrent, higher-order programs.

The language modeled in this machine (Figure~\ref{fig:parallel-anf}) is A-Normal Form lambda calculus~\cite{mattmight:Flanagan:1993:ANF} augmented
with a core set of primitives for multithreaded programming.
\begin{figure}
\begin{align*}
 \expr \in \syn{Exp} &\produces \letiform{\vv}{\call}{\expr}
 \\
 &\;\;\opor\;\; \call
 \\
 &\;\;\opor\;\; \aexpr
 \\
 \call \in \syn{CExp} &\produces \appform{\fexpr}{\aexpr_1 \ldots \aexpr_n}
 \\
 &\;\;\opor\;\; \appform{{\tt callcc}}{\aexpr}
 \\
 &\;\;\opor\;\; \setbform{\vv}{\aexpr_{\mathrm{value}}}
 \\
 &\;\;\opor\;\; \ifform{\aexpr}{\call}{\call}
 \\
 &\;\;\opor\;\; \appform{{\tt cas}}{\vv\; \aexpr_{\mathrm{old}}\; \aexpr_{\mathrm{new}}}
 \\ 
 &\;\;\opor\;\; \appform{{\tt spawn}}{\expr}
 \\
 &\;\;\opor\;\; \appform{{\tt join}}{\aexpr}
\\
 \fexpr, \aexpr \in \syn{AExp} &\produces  \lam \opor \vv \opor n \opor \schfalse 
 \\
 \lam \in \syn{Lam} &\produces \lamform{\vv_1 \ldots \vv_n}{\expr}
 \\
\end{align*}
\caption{ANF lambda-calculus augmented with a core set of primitives for concurrency}
\label{fig:parallel-anf}
\end{figure}
For concurrency, it features an atomic compare-and-swap operation, a spawn form
to create a thread from an expression and a join operation to wait for another
thread to complete.
For higher-order computation, it features closures and first-class continuations.
A closure is a first-class procedure constructed by pairing a lambda term with
an environment that fixes the meaning of its free variables.
A continuation reifies the sequential control-flow for the remainder of the thread as 
a value; when a continuation is ``invoked,'' it restores that control-flow.
Continuations may be invoked an arbitrary number of times, and at 
any time since their moment of creation.

\subsection{\PCEKS{}: A concrete state-space}
A concrete state of execution in the \PCEKS{} machine contains a set of threads plus a shared store.
Each thread is a context combined with a thread id.
A context contains the current expression, 
the current environment,
an address pointing to the current continuation,
and a thread history:
\begin{align*}
\state \in \State &= \s{Threads} \times \s{Store}
\\
\threads \in \s{Threads} &= \Pow{\s{Context} \times \s{TID}}
\\
\ctxt \in \s{Context} &= \syn{Exp} \times \s{Env} \times \s{Addr} \times \s{Hist}
\\
\env \in \s{Env} &= \syn{Var} \parto \s{Addr}
\\
\cont \in \s{Kont} &= \syn{Var} \times \syn{Exp} \times \s{Env} \times \s{Addr} + \set{\mathbf{halt}}
\\
\hist \in \s{Hist} &\text{ contains records of thread history}
\\
\store \in \s{Store} &= \s{Addr} \to \Den 
\\
\den \in \Den &= \s{Value}
\\
\val \in \s{Value} &= \s{Clo} + \s{Bool} + \s{Num} + \s{Kont} + \s{TID} + \s{Addr}
\\
\clo \in \s{Clo} &= \syn{Lam} \times \s{Env} 
\\
\addr \in \s{Addr} &\text{ is an infinite set of addresses}
\\
\tid \in \s{TID} &\text{ is an infinite set of thread ids}
\text.
\end{align*}
The \PCEKS{} machine allocates continuations in the store; thus, to add first-class continuations,
we have first-class addresses.
Under abstraction, program history determines the context-sensitivity of an individual thread.
To allow context-sensitivity to be set as external parameter, we'll leave
program history opaque.  
(For example, to set up a $k$-CFA-like analysis, the program history would be the sequence of calls
made since the start of the program.)
To parameterize the precision of MHP analysis, the thread \IDs{} are also opaque.

\subsection{\PCEKS: A factored transition relation}
Our goal is to factor the semantics of the \PCEKS{} machine, so that one can
drop in a classical CESK machine to model sequential language features.
The abstract interpretation maintains the same
factoring, so that existing analyses of higher-order programs may be ``plugged
into'' the framework for handling concurrency.
The relation $(\To)$ models concurrent transition,
and the relation $(\to)$ models sequential transition:
\begin{align*}
 (\To) &\subseteq \State \times \State
 \\
 (\to) &\subseteq (\s{Context} \times \s{Store}) \times (\s{Context} \times \s{Store})
\end{align*}
For instance, the concurrent transition relation invokes the sequential transition relation
to handle {\tt if}, {\tt set!}, {\tt cas}, {\tt callcc} or procedure call:\footnote{The transition for {\tt cas} is ``sequential'' in the sense that its action is atomic.}
\begin{equation*}
 \infer{
   (\ctxt,\store) \to (\ctxt',\store')
 }{
   (\set{(\ctxt,\tid)} \uplus \threads, \store) \To (\set{(\ctxt',\tid)} \union \threads, \store')
 }
\end{equation*}

Given a program $\expr$, 
the injection function 
$\Inject : \syn{Exp} \to \s{State}$
creates the initial machine state:
\begin{equation*}
  \Inject(\expr)
  =
  ( \set{ ( (\expr, [], \addr_{\mathbf{halt}}, \hist_0), \tid_0 ) } , [\addr_{\mathbf{halt}} \mapsto \mathbf{halt}] )
  \text,
\end{equation*}
where 
$\tid_0$ is the distinguished initial thread \ID,
$\hist_0$ is a blank history
and $\addr_{\mathbf{halt}}$ is the distinguished address of
the $\mathbf{halt}$ continuation.
The meaning of a program $\expr$ is the (possibly infinite) set of states reachable from
the initial state:
\begin{equation*}
  \setbuild{ \state }{ \Inject(\expr) \To^* \state }\text.
\end{equation*}

\paragraph{Sequential transition example: \texttt{callcc}}

There are ample resources dating to Felleisen and Friedman~\cite{mattmight:Felleisen:1986:CEK} detailing the
transition relation of a CESK machine.
For a recent treatment that covers both concrete and abstract transition, see Van Horn and Might~\cite{dvanhorn:VanHorn2010Abstracting}.
Most of the transitions are straightforward, but in the interest
of more self-containment, we review the \texttt{callcc} transition:
\begin{align*}
   ( \overbrace{(\sembr{\appform{{\tt callcc}}{\aexpr}},\env,\addr_\cont,\hist)}^{\ctxt} 
   , 
   \store)
   &\To
   ((\expr, \env'', \addr_\cont,\hist'),\store')
   \text{, where}
   \\
   \hist' &= \record(\ctxt,\hist)
   \\
   (\sembr{\lamform{\vv}{\expr}},\env') &= \Eval(\aexpr,\env,\store)
   \\
   \addr &= \alloc(\vv,\hist')
   \\
   \env'' &= \env'[\vv \mapsto \addr]
   \\
   \store' &= \store[\addr \mapsto \addr_\cont]
   \text.
\end{align*}
The atomic evaluation function $\Eval : \syn{AExp} \times \s{Env} \times \s{Store} \parto \Den$ 
maps an atomic expression to a value in the context of an environment and a store; for example:
\begin{align*}
  \Eval(\vv,\env,\store) &= \store(\env(\vv))
  \\
  \Eval(\lam,\env,\store) &= (\lam, \env)
  \text.
\end{align*}
(The notation $f[x \mapsto y]$ is functional extension:
  the function identical to $f$, except that $x$ now yields $y$ instead of $f(x)$.)

\subsection{A shift in perspective}
Before proceeding, it is worth shifting the formulation so as to ease the process of
abstraction.
For instance, the state-space is well-equipped to handle a finite abstraction over addresses, since we can promote the
range of the store to \emph{sets} of values.
This allows multiple values to live at the same address once an address has been re-allocated.
The state-space is less well-equipped to handle the approximation on thread \IDs{}.
When abstracting thread \IDs{}, we could keep a set of abstract threads paired with the store.
But, it is natural to define the forthcoming concrete and abstract transitions when the set of threads becomes a map.
Since every thread has a distinct thread \ID{}, we can model the set of threads in each state
as a partial map from a thread \ID{} to a context:
\begin{equation*}
 \s{Threads} \equiv \s{TID} \parto \s{Context}
 \text.
\end{equation*}
It is straightforward to update the concurrent transition relation when it
calls out to the sequential transition relation:
\begin{equation*}
 \infer{
   (\ctxt,\store) \to (\ctxt',\store')
 }{
   (\threads[\tid \mapsto \ctxt], \store) \To (\threads[\tid \mapsto \ctxt'], \store')
   \text.
 }
\end{equation*}

\subsection{Concurrent transition in the \PCEKS{} machine}

We define the concurrent transitions separately from the sequential transitions.
For instance, if a context is attempting to {\tt spawn} a thread, the concurrent relation
handles it by allocating a new thread id $\tid'$, and binding it to the new context $\ctxt''$:
\begin{align*}
  (\threads[ \tid \mapsto 
    \overbrace{(\sembr{\appform{{\tt spawn}}{\expr}},\env,\addr_\cont,\hist)}^{\ctxt} ]
    , \store)                                                      
&\To
  (\threads [ \tid \mapsto \ctxt', \tid' \mapsto \ctxt'' ], \store')\text{,}
  \\
  \text{where } \tid' &= \newtid(\ctxt,\threads[\tid \mapsto \ctxt])
  \\
  \ctxt'' &= (\expr,\env,\addr_{\mathbf{halt}},\hist_0)
  \\
  \hist' &= \record(\ctxt,\hist)
  \\
  (\vv', \expr', \env', \addr'_\cont) &= \store(\addr_\cont)
  \\
  \addr' &= \alloc(\vv',\hist') 
  \\
  \env'' &= \env'[\vv' \mapsto \addr']
  \\
  \ctxt' &= (\expr', \env'', \addr'_\cont, \hist') 
  \\
  \store' &= \store[\addr' \mapsto \tid']\text{, where: }
\end{align*}
\begin{itemize}
\item
$\newtid : \s{Context} \times \s{Threads} \to \s{TID}$ 
allocates a fresh thread \ID{} for the newly spawned thread.
\item
$\record : \s{Context} \times \s{Hist} \to \s{Hist}$ is responsible for
updating the history of execution with this context.
\item
$\alloc : \syn{Var} \times \s{Hist} \to \s{Addr}$ allocates a fresh address 
for the supplied variable.
\end{itemize}
The abstract counterparts to these functions determine the degree of approximation in the analysis,
and consequently, the trade-off between speed and precision.

When a thread halts, its thread \ID{} is treated as an address, and its return value is stored there:
\begin{align*}
  (\threads[ \tid \mapsto 
  (\sembr{\aexpr},\env,\addr_{\mathbf{halt}},\hist) ]
    , \store)                                                      
   &\To
  (\threads , \store[\tid \mapsto \Eval(\aexpr,\env,\store)])\text{.}
\end{align*}

This convention, of using thread \IDs{} as addresses, makes it easy to model thread joins, since they can check to see if 
that address has value waiting or not:
\begin{equation*} 
  \infer {
    \store(\Eval(\aexpr,\env,\store)) = \den
  }{
   (\threads[ \tid \mapsto 
    \underbrace{(\sembr{\appform{{\tt join}}{\aexpr}},\env,\addr_\cont,\hist)}_{\ctxt} ]
     , \store)                                                      
     \To
    (\threads[\tid \mapsto (\expr,\env',\addr'_\cont,\hist')] , \store')
    \text,
  }
\end{equation*}
\begin{align*}
 \text{where } \cont &= \store(\addr_\cont)
  \\
  (\vv,\expr,\env,\addr'_\cont) &= \cont
  \\
  \env' &= \env[\vv \mapsto \addr'']
  \\
  \hist' &= \record(\ctxt,\hist)
  \\
  \addr'' &= \alloc(\vv,\hist')
  \\
  \store' &= \store[\addr'' \mapsto \den]
  \text.
\end{align*}

\section{A systematic abstract interpretation of \PCEKS{}}

Using the techniques outlined in our recent work on systematically constructing
abstract interpretations from abstract machines~\cite{dvanhorn:VanHorn2010Abstracting}, we can directly convert the
\PCEKS{} machine into an abstract interpretation of itself.
In the concrete state-space, there are four points at which we must inflict abstraction:
over basic values (like numbers), over histories, over addresses and over thread \IDs{}.

The abstraction over histories  determines the context-sensitivity of the
analysis on a per-thread basis.
The abstraction over addresses determines polyvariance.
The abstraction over thread \IDs{} maps concrete threads into abstract threads,
which determines to what extent the analysis can distinguish dynamically
created threads from one another; it directly impacts MHP analysis.

The abstract state-space (Figure~\ref{fig:abstract-state-space}) 
mirrors the concrete state-space in structure.
\begin{figure}
\begin{align*}
\astate \in \aState &= \sa{Threads} \times \sa{Store}
\\
\athreads \in \sa{Threads} &= \sa{TID} \to \PowSm{\sa{Context}}
\\
\actxt \in \sa{Context} &= \syn{Exp} \times \sa{Env} \times \sa{Addr} \times \sa{Hist}
\\
\aenv \in \sa{Env} &= \syn{Var} \parto \sa{Addr}
\\
\acont \in \sa{Kont} &= \syn{Var} \times \syn{Exp} \times \sa{Env} \times \sa{Addr} + \set{\mathbf{halt}}
\\
\ahist \in \sa{Hist} &\text{ contains bounded, finite program histories}
\\
\astore \in \sa{Store} &= \sa{Addr} \to \aDen 
\\
\aden \in \aDen &= \PowSm{\sa{Value}}
\\
\aval \in \sa{Value} &= \sa{Clo} + \s{Bool} + \sa{Num} + \sa{Kont} + \sa{TID} + \sa{Addr}
\\
\aclo \in \sa{Clo} &= \syn{Lam} \times \sa{Env} 
\\
\aaddr \in \sa{Addr} &\text{ is a finite set of abstract addresses}
\\
\atid \in \sa{TID} &\text{ is a finite set of abstract thread ids}
\end{align*}
\caption{Abstract state-space for a systematic abstraction of the \PCEKS{} machine.}
\label{fig:abstract-state-space}
\end{figure}
We assume the natural point-wise, element-wise and member-wise lifting of a partial order $(\wt)$ over all of the 
sets within the state-space.
Besides the restriction of histories, addresses and thread \IDs{} to finite sets, it is also worth pointing out that
the range of both $\sa{Threads}$ and $\sa{Store}$ are \emph{power} sets.
This promotion occurs because, during the course of an analysis, re-allocating the same thread \ID{} or address is all but inevitable.
To maintain soundness, the analysis must be able to store multiple thread contexts in the same abstract thread \ID{}, and multiple values at the same 
address in the store.

The structural abstraction map $\absmap$ on the state-space (Figure~\ref{fig:abstraction})
utilizes a family of abstraction maps over the sets within the state-space.
\begin{figure}
\begin{align*}
\absmap_\State(\threads,\store) 
   &= (\absmap(\threads), \absmap(\store))
\\
\absmap_{\s{Threads}}(\threads)
   &= \lambda \atid .
   \!\! \bigsqcup_{\absmap(\tid) = \atid} \!\!
   \absmap(\threads(\tid))
\\
\absmap_{\s{Context}}(\expr,\env,\cont,\hist)
   &= \set{ (\expr,\absmap(\env), \absmap(\cont), \absmap(\hist)) }
\\
\absmap_{\s{\Env}} (\env)
   &= \lambda \vv . \absmap(\env(\vv))
\\
\absmap_{\s{Kont}} (\vv,\expr,\env,\addr) 
   &= (\vv,\expr,\absmap(\env), \absmap(\addr))
\\
\absmap_{\s{Kont}} (\mathbf{halt})
   &= \mathbf{halt}
\\
\absmap_{\s{Store}} (\store)
   &= \lambda \aaddr .
   \!\! \bigsqcup_{\absmap(\addr) = \aaddr} \!\!
   \absmap(\store(\addr))
\\
\absmap_{\Den} (\val) &= \set{ \absmap(\val) }
\\
\absmap_{\s{Clo}} (\lam, \env) &= (\lam, \absmap(\env))
\\
\absmap_{\s{Bool}} (b) &= b
\\
\absmap_{\s{Hist}} (\hist) &\text{ is defined by context-sensitivity}
\\
\absmap_{\s{TID}} (\tid) &\text{ is defined by thread-sensitivity}
\\
\absmap_{\s{Addr}} (\addr) &\text{ is defined by polyvariance}
\text.
\end{align*}
\caption{A structural abstraction map.}
\label{fig:abstraction}
\end{figure}
With the abstraction and the abstract state-space fixed,
the abstract transition relation reduces to a matter of calculation~\cite{mattmight:Cousot:1979:Galois}.
The relation $(\aTo)$ describes the concurrent abstract transition, while the relation $(\ato)$ describes the
sequential abstract transition:
\begin{align*}
 (\aTo) &\subseteq \aState \times \aState
 \\
 (\ato) &\subseteq (\sa{Context} \times \sa{Store}) \times (\sa{Context} \times \sa{Store})
\end{align*}
When the context in focus is sequential, the sequential relation takes over:
\begin{equation*}
 \infer{
   (\actxt,\astore) \ato
   (\actxt',\astore')
 }{
   (\athreads[\atid \mapsto \set{\actxt} \union \hat{C}], \astore) \aTo
   (\athreads \join [\atid \mapsto \set{\actxt'}], \astore')
 }
\end{equation*}
There is a critical change over the concrete rule in this abstract rule: thanks to the join operation, the abstract context remains associated
with the abstract thread \ID{} even after its transition has been considered.
In the next section, we will examine the application of singleton abstraction to thread \IDs{} to allow the ``strong update'' of 
abstract threads \IDs{} across transition.
(For programs whose maximum number of threads is statically bounded by a known constant, this allows for precise MHP analysis.)

\subsection{Running the analysis}
Given a program $\expr$, 
the injection function $\aInject : \syn{Exp} \to \sa{State}$
 creates the initial abstract machine state:
\begin{equation*}
  \aInject(\expr)
  =
  \left( \left[\atid_0 \mapsto \set{ (\expr, [], \aaddr_{\mathbf{halt}}, \ahist_0) } \right] , 
    [\aaddr_{\mathbf{halt}} \mapsto \set{ \mathbf{halt} } ] \right)
  \text,
\end{equation*}
where $\ahist_0$ is a blank abstract history
and $\aaddr_{\mathbf{halt}}$ is the distinguished abstract address of
the $\mathbf{halt}$ continuation.
The analysis of a program $\expr$ is the finite set of states reachable from
the initial state:
\begin{equation*}
  \aReachable(\expr) = \setbuild{ \astate }{ \aInject(\expr) \aTo^* \astate }\text.
\end{equation*}

\subsection{A fixed point interpretation}
If one prefers a traditional, fixed-point abstract interpretation,
we can imagine the intermediate state of the analysis itself as a set
of currently reachable abstract machine states:
\begin{equation*}
  \amstate \in \aMState = \PowSm{\aState}
  \text.
\end{equation*}
A global transfer function $\atf : \aMState \to \aMState$ evolves this set:
\begin{align*}
 \atf(\amstate) &= \set{ \aInject(\expr) } \union \setbuild{ \astate' }{ \astate \in \amstate \text{ and } \astate \aTo \astate' }
 \text.
\end{align*}
The solution of the analysis is the least fixed point: $\lfp(\atf)$.

\subsection{Termination}
The dependence structure of the abstract state-space is a directed acyclic
graph starting from the set $\aState$ at the root.
Because all of the leaves of this graph (\eg, lambda terms, abstract numbers,
abstract addresses) are finite for any given program, the state-space itself
must also be finite.
Consequently, there are no infinitely ascending chains in the lattice
$\aMState$.
By Kleene's fixed point theorem, there must exist a least natural $n$ such
that $\lfp(\atf) = \atf^n(\emptyset)$.

\subsection{Concurrent abstract transitions}

Guided by the structural abstraction, we can convert the concrete concurrent transitions for the \PCEKS{} machine
into concurrent abstract transitions.
For instance, if an abstract context is attempting to {\tt spawn} a thread, the concurrent relation
handles it by allocating a new thread id $\atid'$, and binding it to the new context $\actxt''$:
\begin{align*}
  (\athreads [ \atid \mapsto 
    \setsm{ \overbrace{(\sembr{\appform{{\tt spawn}}{\expr}},\aenv,\aaddr_\acont,\ahist)}^{\actxt} }
    \union \hat C
  ]
    , \astore ) 
&\aTo
  (\athreads \join [ \atid \mapsto \set{\actxt'}, \atid' \mapsto \set{\actxt''} ], \store')\text{,}
  \\
  \text{where } \atid' &= \anewtid(\actxt,\athreads[\atid \mapsto \hat C \union \set{\actxt}])
  \\
  \actxt'' &= (\expr,\aenv,\aaddr_{\mathbf{halt}},\ahist_0)
  \\
  \ahist' &= \arecord(\actxt,\ahist)
  \\
  (\vv', \expr', \aenv', \aaddr'_\acont) &\in \astore(\aaddr_\acont)
  \\
  \aaddr' &= \aalloc(\vv',\ahist') 
  \\
  \aenv'' &= \aenv'[\vv' \mapsto \aaddr']
  \\
  \actxt' &= (\expr', \aenv'', \aaddr'_\acont, \ahist') 
  \\
  \astore' &= \astore \join [\aaddr' \mapsto \set{\atid'}]\text{, where: }
\end{align*}
\begin{itemize}
\item
$\anewtid : \sa{Context} \times \sa{Threads} \to \sa{TID}$ 
allocates a thread \ID{} for the newly spawned thread.
\item
$\arecord : \sa{Context} \times \sa{Hist} \to \sa{Hist}$ is responsible for
updating the (bounded) history of execution with this context.
\item
$\aalloc : \syn{Var} \times \sa{Hist} \to \sa{Addr}$ allocates an address for
 the supplied variable.
\end{itemize}
These functions determine the degree of approximation in the analysis,
and consequently, the trade-off between speed and precision.

When a thread halts, its abstract thread \ID{} is treated as an address, and its
return value is stored there: 
\begin{equation*}
\infer{ \athreads' = \athreads \join [ 
     \atid \mapsto \setsm{ (\sembr{\aexpr},\aenv,\aaddr_{\mathbf{halt}},\ahist) } ] }
{ ( \athreads' , \store)
  \aTo (\athreads' , \astore \join [\atid \mapsto
  \aEval(\aexpr,\aenv,\astore)])
}  \text{, where }
\end{equation*}
the atomic evaluation function $\aEval : \syn{AExp} \times \sa{Env} \times \sa{Store} \to \aDen$ 
maps an atomic expression to a value in the context of an environment and a store:
\begin{align*}
  \aEval(\vv,\aenv,\astore) &= \astore(\aenv(\vv))
  \\
  \aEval(\lam,\aenv,\astore) &= \setsm{ (\lam, \aenv) }
  \text.
\end{align*}
It is worth asking whether it is sound in just this case to remove the context
from the threads (making the subsequent threads $\threads$ instead of $\threads'$).
It is sound, but it seems to require a (slightly) more complicated staggered-state bisimulation to prove it: the
concrete counterpart to this state may take several steps to eliminate all of
its halting contexts.

Thanks to the convention to use thread \IDs{} as
addresses holding the return value of  thread, it easy to model thread joins, since they can check to see if 
that address has a value waiting or not:
\begin{equation*} 
  \infer {
    \aaddr \in \aEval(\aexpr,\aenv,\astore)
    \longand
    \aden = \astore(\aaddr) 
  }{
   (\athreads \join [ \atid \mapsto 
    \setsm{
    \underbrace{(\sembr{\appform{{\tt join}}{\aexpr}},\aenv,\aaddr_\acont,\ahist)}_{\actxt} } ]
     , \astore) 
     \aTo
    (\athreads \join [\atid \mapsto \setsm{(\expr,\aenv',\aaddr'_\acont,\ahist'), \actxt} ] , \astore')
    \text,
  }
\end{equation*}
\begin{align*}
 \text{where } \acont &\in \astore(\aaddr_\acont)
  \\
  (\vv,\expr,\aenv,\aaddr'_\acont) &= \acont
  \\
  \aenv' &= \aenv[\vv \mapsto \aaddr'']
  \\
  \ahist' &= \arecord(\actxt,\ahist)
  \\
  \aaddr'' &= \aalloc(\vv,\ahist')
  \\
  \astore' &= \astore \join [\aaddr'' \mapsto \aden]
  \text.
\end{align*}

\subsection{Soundness}
Compared to a standard proof of soundness for a small-step abstract
interpretation, the proof of soundness requires only slightly more attention 
in a concurrent setting.
The key lemma in the inductive proof of simulation states that when a concrete state $\state$ abstracts to an abstract state $\astate$,
 if the concrete state can transition to $\state'$, then the abstract state $\astate$ must be able to transition to some other abstract state $\astate'$
such that $\state'$ abstracts to $\astate'$:
\begin{theorem}
If: 
\begin{equation*}
 \absmap(\state) \wt \astate \text{ and } 
 \state \To \state'\text,
\end{equation*}
then there must exist a state $\astate'$ such that:
\begin{equation*}
 \absmap(\state') \wt \astate' \text { and } 
 \astate \aTo \astate'\text.
\end{equation*}
\end{theorem}
\begin{proof}
The proof is follows the case-wise structure of proofs like those in Might and Shivers~\cite{mattmight:Might:2008:Exploiting}.
There is an additional preliminary step: first choose a thread id modified across transition, and then 
perform case-wise analysis on how it could have been modified.
\end{proof}

\subsection{Extracting flow information}

The core question in flow analysis is, ``Can the value $\val$ flow to the expression $\aexpr$?'' 
To answer it, assume that $\amstate$ is the set of all reachable abstract states.
We must check every state within this set, and every environment $\aenv$ within that state.
If $\astore$ is the store in that state, then $\val$ may flow to $\aexpr$ if
the value $\absmap(\val)$ is represented in the set $\aEval(\aexpr,\aenv,\astore)$.
Formally, we can construct a flows-to relation, $\mathit{FlowsTo} \subseteq \s{Value} \times \syn{AExpr}$,
for a program $\expr$:
\begin{align*}
  \mathit{FlowsTo}(\val,\aexpr)
  \text{ iff there exist }
  (\athreads,\astore) \in \aReachable(\expr)
  \text{ and }
  \atid \in \dom(\athreads) 
  \text{ such that }
  \\
  (\expr,\aenv,\acont) \in \athreads(\atid) 
  \text{ and }
  \set{\absmap(\val)} \wt \aEval(\aexpr,\aenv,\astore)
\text.
\end{align*}

\subsection{Extracting MHP information}
In MHP analysis, we are concerned with whether two expressions $\expr'$ and $\expr''$
may be evaluated concurrently with one another in program $\expr$.
It is straightforward to decide this using the set of reachable states computed
by the abstract interpretation.
If, in any reachable state, there exist two distinct contexts at the relevant expressions,
then their evaluation may happen in parallel with one another.
Formally, the $\mathit{MHP} \subseteq \syn{Exp} \times \syn{Exp}$ relation
with respect to program $\expr$ is:
\begin{align*}
  \mathit{MHP}(\expr', \expr'') 
  \text{ iff there exist } 
  (\athreads,\astore) \in \aReachable(\expr)
  \text{ and }
  \atid',\atid'' \in \dom(\athreads)
  \text{ such that }
  \\
  (\expr',\_,\_,\_) \in \athreads(\atid') \text{ and }
  (\expr'',\_,\_,\_) \in \athreads(\atid'')
  \text.
\end{align*}

\section{MHP: Making strong transitions with singleton threads}

In the previous section, we constructed a systematic abstraction of the \PCEKS{} machine.
While it serves as a sound and capable flow analysis, 
its precision as an MHP analysis 
is just above useless.
Contexts associated with each abstract thread \ID{} grow monotonically
during the course of the analysis.
Eventually, it will seem as though every context may happen in parallel with every other context.
By comparing the concrete and abstract semantics, the cause of the imprecision
becomes clear:
where the concrete semantics \emph{replaces} the context at a given thread \ID{},
the abstract semantics \emph{joins}.

Unfortunately, we cannot simply discard the join.
A given abstract thread \ID{} could be representing multiple concrete thread \IDs{}.
Discarding a thread \ID{} would then discard possible interleavings, and it could even introduce unsoundness.

Yet, it is plainly the case that \emph{many} programs have 
a boundable number of threads that are co-live.
Thread creation is considered expensive, and thread pools created during program initialization are a popular mechanism 
for circumventing the problem.
To exploit this design pattern, we can make thread \IDs{} eligible for ``strong update'' across transition.
In shape analysis, strong update refers to the ability to treat an abstract address as the
representative of a single concrete address when assigning to that address.
That is, by  tracking the cardinality of the abstraction of each thread \ID{}, we can determine when it is sound
to replace functional join with functional update \emph{on threads themselves}.

The necessary machinery is straightforward, adapted directly from the shape analysis literature~\cite{mattmight:Balakrishnan:2006:Recency,mattmight:Chase:1990:Analysis,mattmight:Jagannathan:1998:Single,mattmight:Hudak:1985:Sharing,mattmight:Might:2008:Exploiting,mattmight:Sagiv:2002:TVLA} ;
we attach to each state a cardinality counter $\atcount$ that tracks how many times an abstract thread \ID{} has been allocated (but not precisely beyond once):
\begin{align*}
\astate \in \aState &= \sa{Threads} \times \sa{Store} \times \sa{TCount}
\\
\atcount \in \sa{TCount} &= \sa{TID} \to \set{0,1,\infty}
\text.
\end{align*}
When the count of an abstract thread \ID{} is exactly one, we know for certain that there exists at most one concrete counterpart.
Consequently, it is safe to perform a ``strong transition.''
Consider the case where the context in focus for the concurrent transition is sequential; 
in the case where the count is exactly one, the abstract context gets replaced on transition:
\begin{equation*}
 \infer{
   (\actxt,\astore) \ato
   (\actxt',\astore')
   \longand
   \atcount(\atid) = 1
 }{
   (\athreads[\atid \mapsto \set{\actxt} \uplus \hat{C}], \astore,\atcount) \aTo
   (\athreads[\atid \mapsto \set{\actxt'} \union \hat{C}], \astore',\atcount)
   \text.
 }
\end{equation*}
It is straightforward to modify the existing concurrent transition rules to exploit information available in the cardinality counter.
At the beginning of the analysis, all abstract thread \IDs{} have a count of zero.
Upon spawning a thread, the analysis increments the result of the function $\anewtid$.
When a thread whose abstract thread \ID{} has a count of one halts, its count is reset to zero.

\subsection{Strategies for abstract thread \ID{} allocation}
Just as the introduction of an allocation function for addresses provides the ability to tune
polyvariance, the $\anewtid$ function provides the ability to tune precision.
The optimal strategy for allocating this scarce pool of abstract thread \IDs{} depends upon the design patterns in use.

One could, for instance, allocate abstract thread \IDs{} according to calling context, \eg, the abstract thread \ID{} is the last $k$ call sites.
This strategy would work well for the implementation of futures, where futures from the same context are often not
co-live with themselves.

The context-based strategy, however, is not a reasonable strategy for a
thread-pool design pattern.
All of the spawns will occur at the same expression in the same loop, and
therefore, in the same context.
Consequently, there will be no discrimination between threads.
If the number of threads in the pool is known \emph{a priori} to be $n$, then
the right strategy for this pattern is to create $n$ abstract thread \IDs{}, and
to allocate a new one for each iteration of the thread-pool-spawning loop.
On the other hand, if the number of threads is set dynamically, no amount of abstract thread \IDs{} will 
be able to discriminate between possible interleavings effectively, 
in this case a reasonable choice for precision would be to have one abstract thread per thread pool.

\subsection{Advantages for MHP analysis}
With the cardinality counter, it is possible to test whether an expression may be evaluated in parallel with itself.
If, for every state, every abstract thread id which maps to a context containing that expression has a count of one,
and no other context contains that expression, then that expression must never be evaluated in parallel with itself.
Otherwise, parallel evaluation is possible.

\section{Flow analysis of concurrent higher-order programs}

If the concern is a sound flow analysis, but not MHP analysis, then we can
perform an abstract interpretation of our abstract interpretation that
efficiently collapses all possible interleavings and paths, even as it
retains limited (reachability-based) flow-sensitivity.
This second abstraction map $\absmap' : \aMState \to \aState$ operates on the system-space of the fixed-point interpretation:
\begin{align*}
\absmap'(\amstate) &= 
\bigsqcup_{\astate \in \amstate} \astate
\text.
\end{align*}
The new transfer function, $\atf' : \aState \to \aState$, 
monotonically accumulates all of the visited states into a single state:
\begin{equation*}
 \atf'(\astate) = \astate \join 
  \bigsqcup_{\astate \To \astate'} \astate'
 \text.
\end{equation*}

\subsection{Complexity}
This second abstraction simplifies
the calculation of an upper bound on computational complexity.
The structure of the set $\aState$ is a pair of maps into sets:
\begin{equation*}
 \aState = 
 \left( \sa{TID} \to \PowSm{\sa{Context}}  \right)
 \times
 \left( 
  \sa{Addr} \to \PowSm{\sa{Value}}
 \right)
 \text.
\end{equation*}
Each of these maps is, in effect, a table of bit vectors:
the first with abstract thread \IDs{} on one axis and contexts on the other; 
and the second with abstract addresses on one axis and values on the other.
The analysis monotonically flips bits on each pass.
Thus, the maximum number of passes---the tallest ascending chain in the lattice $\aState$---is:
\begin{equation*}
  \abs{\sa{TID}} \times \abs{\sa{Context}}
  +
  \abs{\sa{Addr}} \times \abs{\sa{Value}}
  \text.
\end{equation*}
Thus, the complexity of the analysis is 
determined by context-sensitivity, as with classical sequential flow analysis.
For a standard monovariant analysis, the complexity is polynomial~\cite{mattmight:Palsberg:1995:CFA}.
For a context-sensitive analysis with shared environments, the complexity is exponential~\cite{dvanhorn:VanHorn-Mairson:ICFP08}.
For a context-sensitive analysis with flat environments, the complexity is again polynomial~\cite{dvanhorn:Might2010Resolving}.

\section{Related work}

This work traces its ancestry to Cousot and Cousot's work on abstract interpretation~\cite{mattmight:Cousot:1977:AI,mattmight:Cousot:1979:Galois}.
We could easily extend the fixed-point formulation with the implicit concretization function to arrive at
an instance of traditional abstract interpretation.
It is also a direct descendant of the line of work investigating control-flow in higher-programs
that began with Jones~\cite{mattmight:Jones:1981:LambdaFlow} and Shivers~\cite{dvanhorn:shivers-88,mattmight:Shivers:1991:CFA}.

The literature on static analysis of concurrency and higher-orderness is not empty, but it is spare.
Much of it focuses on the special case of the analysis of futures.
The work most notable and related to our own is that of Navabi and Jagannathan~\cite{mattmight:Navabi:2009:Futures}.
It takes Flanagan and Felleisen's notion of safe futures~\cite{mattmight:Flanagan:1995:Future,mattmight:Flanagan:1999:Futures},
and develops a dynamic and static analysis that can prevent a continuation from modifying a resource 
that one of its concurrent futures may modify.
What makes this work most related to our own is that it is sound even in the presence of exceptions, which are, in essence,
an  upward-restricted form of continuations.
Their work and our own own interact synergistically, since their safety analysis focuses on removing the parallel inefficiencies
of safe futures; our flow analysis can remove the sequential inefficiencies of futures through the elimination of run-time
type-checks.
Yahav's work is the earliest to apply shape-analytic techniques to the
analysis of concurrency~\cite{mattmight:Yahav:2001:Verifying}.

It has taken substantial effort to bring the static analysis of
higher-order programs to heel; to recite a few of the major
challenges:
\begin{enumerate} 

\item First-class functions from dynamically created closures
over lambda terms create recursive dependencies between control- and data-flow;
$k$-CFA co-analyzes control and data to factor and order these dependencies~\cite{mattmight:Shivers:1991:CFA}.

\item Environment-bearing closures over lambda terms impart fundamental
intractabilities unto context-sensitive
analysis~\cite{dvanhorn:VanHorn-Mairson:ICFP08}---intractabilities that were only recently
side-stepped via flattened abstract environments~\cite{dvanhorn:Might2010Resolving}.
\item The functional emphasis on recursion over iteration made achieving high
precision difficult (or hopeless) without abstract garbage
collection to recycle tail-call-bound parameters and continuations~\cite{mattmight:Might:2008:Exploiting}.
\item When closures keep multiple bindings to the same variable live, precise reasoning about side effects to these bindings requires the
adaptation of shape-analytic techniques~\cite{mattmight:Jagannathan:1998:Single,mattmight:Might:2010:Shape}.
\item Precise reasoning about first-class continuations (and kin such as
exceptions) required a harmful conversion to continuation-passing style until the advent
of small-step abstraction interpretations for the pointer-refined CESK
machine~\cite{dvanhorn:VanHorn2010Abstracting}.  

\end{enumerate}
We see this work as another milestone on the path to
robust static analysis of full-featured higher-order programs.

\section{Limitations and future work}
Since the shape of the store and the values within were not the primary focus
of this work, it utilized a blunt abstraction.
A compelling next step of this work would generalize the abstraction of the
store relationally, so as to capture relations between the values at specific
\emph{addresses}.
The key challenge in such an extension is the need to handle relations
between abstract addresses which may represent \emph{multiple} concrete
addresses.
Relations which universally quantify over the concrete
constituents of an abstract address are a promising approach.

\paragraph{Acknowledgements}

 This material is based upon work supported by the NSF under Grant No. 1035658.
 The second author was supported by the National Science Foundation under Grant
 No. 0937060 to the Computing Research Association for the CIFellow Project.
 We also acknowledge support from the Park City Research Institute for
 Computer Science.

%


\bibliographystyle{acm}
\bibliography{bibliography}

\end{document}